\documentclass[prb,twocolumn,showpacs,byrevtex,showkeys]{revtex4}

\usepackage{graphics}
\usepackage{dcolumn}
\usepackage{amsmath}
\usepackage{amssymb}

\newcommand{\ie}{{\it i.e.}}
\newcommand{\eg}{{\it e.g.}}

\newcommand{\vs}{{\it vs.}}
\newcommand{\etc}{{\it etc.}}

\newcommand{\etal}{{\it et~al.}}

\newcommand{\HS}{\ensuremath{{\cal H}}}
\newcommand{\G}{\ensuremath{{\cal G}}}
\newcommand{\YBCO}{YBa$_2$Cu$_3$O$_7$}

\newcommand{\FloatPlacement}{!tb}

\newcommand{\state}[1]{$\stat{#1}$}

\newcommand{\stat}[1]{%
  \begingroup
  \def\delimiter##1##2##3##4##5##6##7##8{##10##3##4##5}%
  \mathcode`\0=\downarrow \mathcode`\1=\uparrow
  \mathcode`\d=\downarrow \mathcode`\u=\uparrow
  \mathcode`\D=\Downarrow \mathcode`\U=\Uparrow
  \mathcode`\-=\circ
  #1\relax
  \endgroup
}

\begin{document}

\title{
  Ground state and bias current induced rearrangement of semifluxons in 0-$\pi$ long Josephson junctions.
}

\author{E.~Goldobin}
\email{gold@uni-tuebingen.de}
\homepage{http://www.geocities.com/e_goldobin}
\author{D.~Koelle}
\author{R.~Kleiner}
\affiliation{
  Physikalisches Institut - Experimentalphysik II,
  Universit\"at T\"ubingen,
  Auf der Morgenstelle 14,
  D-72076 T\"ubingen, Germany
}

\pacs{
  74.50.+r,   
  85.25.Cp    
  74.20.Rp    
}

\keywords{
  Long Josephson junction, sine-Gordon,
  half-integer flux quantum, semifluxon, 
  0-pi-junction
}

\date{\today}

\begin{abstract}
  We investigate numerically a long Josephson junction with several phase $\pi$-discontinuity points. Such junctions are usually fabricated as a ramp between an anisotropic cuprate superconductor like {\YBCO} and an isotropic metal superconductor like Nb. From the top, they look like zigzags with $\pi$-jumps of the Josephson phase at the corners. These $\pi$-jumps, at certain conditions, lead to the formation of half-integer flux quanta, which we call semifluxons (SF), pinned at the corners. We show (a) that the spontaneous formation of SFs depends on the junction length, (b) that the ground state without SFs can be converted to a state with SFs by applying a bias current, (c) that the SF configuration can be rearranged by the bias current. All these effects can be observed using a SQUID microscope.
  
\end{abstract}

\maketitle

\section{Introduction}
\label{Sec:Intro}

Due to the specific order parameter symmetry of high-$T_c$ cuprate superconductors it is possible to fabricate and study so-called $0$ and $\pi$ Josephson junctions (see Ref.~\onlinecite{Tsuei:Review} and references therein). These junctions can be fabricated in various geometries and open new perspectives for Josephson electronics (digital circuits, fluxon devices, quantum bits, \etc)\cite{Tsuei:d-wave:implications}. 

Recent experiments\cite{Smilde:ZigzagPRL} with {\YBCO}-Nb ramp long Josephson junctions (LJJ) fabricated in zigzag geometry (if viewed from the top) clearly demonstrated that such a LJJ consists of alternating facets of $0$, $\pi$, $0$, $\pi\ldots$ junctions. The presence of alternating $0$ and $\pi$-facets results in a set of $\pi$-discontinuities of the Josephson phase at the corners where $0$- and $\pi$-facets join. These are  the points where the order parameter of the anisotropic high-$T_c$ superconductor changes its sign because the direction of the Josephson contact changes by $90^\circ$. 


In the \emph{short facet limit} the $I_c(H)$ dependence of the zigzag LJJ was investigated theoretically and experimentally\cite{Smilde:ZigzagPRL}.   Due to alternating $0$ and $\pi$ facets the critical current $I_c$ has a rather small value at zero magnetic field $H$, while the main peaks of $I_c$ are found at a finite field, the expression for which was derived too and fits well with numerical simulations and experiments (considering flux focusing). 

A striking property of 0-$\pi$ LJJ is the spontaneous generation of half-integer flux quanta, further called semifluxons (SFs), at the corners of a zigzag. The presence of SFs was demonstrated experimentally\cite{Hilgenkamp:zigzag:SF} by scanning SQUID microscopy on LJJs in the \emph{long facet limit}. SFs were also experimentally observed in the so-called tri-crystal grain boundary LJJs
\cite{Kirtley:SF:HTSGB,Kirtley:SF:T-dep,Sugimoto:TriCrystal:SF}. 
In both experiments the samples were electrically disconnected. 


In this work we investigate numerically the behavior of positive and negative SFs (PSF and NSF) in a LJJ of zigzag geometry. The polarity of SFs is defined so that the PSF roughly corresponds to spontaneous magnetic flux $+\Phi_0/2$, while NSF corresponds to $-\Phi_0/2$. 
In particular, we study the various ground states in which SFs of different polarity can be arranged in LJJs with many discontinuity points. We also investigate the effect of the bias current, and show that it may provoke formation of SFs  even in the cases when the ground state has a piecewise constant phase at zero bias. The rearrangement of the PSFs and NSFs under the effect of uniform bias current at zero voltage is discovered and may be observed in experiments similar to the one of Ref.~\onlinecite{Hilgenkamp:zigzag:SF}. Throughout the paper we consider separately the cases of infinitely long JJ and of LJJ of finite length. Among different ground states we focus on antiferromagnetically ordered chains of SFs (at zero bias) and on chains of unipolar SFs. We do not limit ourselves to only one corner but consider several of them, usually equidistantly spaced.

In section \ref{Sec:Model} we introduce the model and discuss numerical aspects. Various ground states are discussed in section \ref{Sec:GroundState}. Section \ref{Sec:Rearrangement} contains numerical results which show the rearrangement of SFs by means of an applied bias current. Section \ref{Sec:Conclusion} concludes this work.

\section{Model and Numerics}
\label{Sec:Model}

The dynamics of the Josephson phase in LJJ with phase $\pi$-discontinuities is described by the following perturbed sine-Gordon equation\cite{Goldobin:LJJwPiPts}:
\begin{equation}
  \phi_{xx} - \phi_{tt} - \sin(\phi)
  =\alpha\phi_t - \gamma(x)
  + h_x(x) + \theta_{xx}(x)
  , \label{Eq:Basic:Norm}
\end{equation}
where $\phi(x,t)$ denotes the Josephson phase across the junction, $\alpha\equiv 1/\sqrt{\beta_c}$ is the dimensionless damping coefficient ($\beta_c$ is the McCumber-Stewart parameter\cite{Likharev:book}). The function $\theta(x)$ describes the positions of the $\pi$-discontinuity points and can be written as
\begin{equation}
  \theta(x) = \pi \sum_{k=1}^{N_c} \sigma_k \HS(x-x_k)
  , \label{Eq:theta}
\end{equation}
where $\sigma_k=\pm 1$ defines the direction of the $k$-th phase jump. The sum is over all $N_c$ corners located at $x=x_k$. $\HS(x)$ is the Heaviside step function. 

Eq.~(\ref{Eq:Basic:Norm}) is written using standard normalized units, \ie, the coordinate and the time are given in units of the Josephson penetration depth $\lambda_J$, and the inverse plasma frequency $\omega_p^{-1}$, respectively\cite{Likharev:book}. The subscripts $x$ and $t$ in Eq.~(\ref{Eq:Basic:Norm}) and below, if any, denote partial derivatives with respect to $x$ and $t$.

The external magnetic field enters into consideration through the $h_x$ term in Eq.~(\ref{Eq:Basic:Norm}) and through the boundary conditions, which for overlap geometry can be written as
\begin{equation}
  \phi_x(0,L) = h
  , \label{Eq:BC}
\end{equation}
where the field $h$ is normalized in the usual way as
\begin{equation}
  h = \frac{2H}{H_{c1}}
  . \label{Eq:FieldNormalization}
\end{equation}
Here, $H_{c1}=\Phi_0/(\pi\mu_0\Lambda\lambda_J)$ is the first critical field (penetration field) for a LJJ which is, in fact, equal to the field in the center of the fluxon. $\Phi_0$ is the quantum of magnetic flux and $\Lambda$ is the magnetic thickness of the junction. The field $H$ in Eq.~(\ref{Eq:FieldNormalization}) is given as $H=(\mathbf{H \cdot n})$, where $\mathbf{n}$ is the unit vector normal to the effective cross-section of the junction (elementary cell). Note, that if the projection of the external field on all facets is the same, the term $h_x(x)$ disappears from Eq.~(\ref{Eq:Basic:Norm}), and the field affects the LJJ only via the boundary condition (\ref{Eq:BC}).

In numerical simulations, it is quite difficult to deal with derivatives of $\delta$-functions present within the $\theta_{xx}$ term in Eq.~(\ref{Eq:Basic:Norm}). Therefore, it is convenient to present the total phase $\phi$ as a sum of two components: a magnetic component $\mu(x)$ and the order-parameter related one $\theta(x)$ (\ref{Eq:theta}):
\begin{equation}
  \phi(x,t) = \mu(x,t) + \theta(x)
  . \label{Eq:}
\end{equation}
In this case we can get the sine-Gordon equation only for the magnetic component of the phase $\mu$:
\begin{equation}
  \mu_{xx} - \mu_{tt} - \sin(\mu)\underbrace{\cos(\theta)}_{\pm1}
  =\alpha\mu_t - \gamma(x) + h_x(x)
  . \label{Eq:Final:mu:Norm}
\end{equation}
This is the usual perturbed sine-Gordon equation, but the sign in front of the sine function changes from facet to facet. We use Eq.~(\ref{Eq:Final:mu:Norm}) for numerical simulations.

Xu {\etal}\cite{Xu:SF-shape,Goldobin:TwinBoundaryComment} and later, but independently, Goldobin {\etal} \cite{Goldobin:LJJwPiPts} derived an analytical expression for the SF in an infinitely long JJ with a single phase $\pi$-discontinuity point at $x=0$
\begin{equation}
  \mu(x) = \left\{
    \begin{array}{lll}
            &4 \arctan\left( {\cal G} e^{x} \right) &;\quad x<0\\
      \pi - &4 \arctan\left( {\cal G}e^{-x} \right) &;\quad x>0
    \end{array}
  \right.
  , \label{Eq:SF}
\end{equation}
where ${\cal G}=\tan(\pi/8)=\sqrt{2}-1$.
The corresponding expression for the magnetic field of a SF is
\begin{equation}
  \mu_x(x) = \frac{2}{\cosh(|x|-\ln{\cal G})}
  . \label{Eq:mu_x}
\end{equation}
It looks like a cusp with exponential tails.

The simulations were done using \textsc{StkJJ} software\cite{StkJJ} further developed to include discontinuity points in the numerical scheme and new initial conditions with arbitrary distributed positive and negative SFs given by Eq.~(\ref{Eq:SF}). For the simulations we use a linear LJJ with a damping parameter $\alpha=0.1$, which corresponds well to the $\alpha$ values between $0.1$ and $1.0$ in real zigzag LJJs\cite{Goldobin:0Pi:AlphaEstimation}. The applied magnetic field was zero.

\section{Ground state}
\label{Sec:GroundState}

To understand the formation of ground states in various configurations it is important to be able to calculate the energy of the system. Since the sine-Gordon Eq.~(\ref{Eq:Final:mu:Norm}) differs from the usual equation only by a factor in front of $\sin\mu$, the energy of the system in the general case can be written as
\begin{equation}
  U
  =\int_{-\infty}^{+\infty} 
  \frac{1}{2}\mu_x^2 + (1-\cos\mu\,\cos\theta)\,dx
  , \label{Eq:U}
\end{equation}
Now we present the results on the observation of different ground states and the spontaneous formation of SFs.

\subsection{Infinitely long JJ}

When we speak about infinitely long JJ we mean that its both edges are very far from the positions of discontinuity points, fluxons, SFs and any other ``interesting'' locations. For our purposes we used $L=20$ ($0\leq x \leq L$) as a good approximation provided that SFs are at least $5\lambda_J$ apart from the edges. For some cases we repeated the simulations using $L=30$ and $L=50$, but the results were \emph{quantitatively} the same with accuracy of few percents. 

First, we simulated a LJJ with {\bf one discontinuity} point at $x_1=10$  and observed that the initial phase distribution 
\begin{equation}
  \phi(x)=\theta(x) = \pi\HS(x-x_1)
  , \label{Eq:theta1}
\end{equation}
is not stable and evolves into a NSF. If we take $\theta(x) = -\pi\HS(x-x_1)$ the system generates a PSF instead of NSF. Note, that in principle the polarity of SF does not depend on the sign of the $\pi$-discontinuity. Both PSF and NSF may be pinned at a $\theta(x)$ or a $-\theta(x)$ discontinuity (\ref{Eq:theta1}) and have the same energy. The energy of the SF, calculated by substituting Eq.~(\ref{Eq:SF}) into Eq.~(\ref{Eq:U}) is
\begin{equation}
  U = 16\frac{\G^2}{1+\G^2}
  , \label{Eq:U_SF}
\end{equation}
while the energy of the $\mu=0$ state is $\propto L$, \ie{} it diverges.
Note that Eqs.~(\ref{Eq:SF}) and (\ref{Eq:U_SF}) are universal for both fluxons and SFs. If we take $\G=1$, Eq.~(\ref{Eq:SF}) converts into the sine-Gordon kink and Eq.~(\ref{Eq:U}) gives $U=8$ -- the rest energy of a fluxon. If we take $\G=\sqrt{2}-1$ (as for a SF), we get the SF shape from Eq.~(\ref{Eq:SF}) and $U=8-4\sqrt{2}\approx 2.343$ from Eq.~(\ref{Eq:U_SF}). This is the rest energy of a SF.

Second, we simulated a LJJ with {\bf two discontinuity points} at $x_1=5$ and $x_2=15$. The initial phase distribution was chosen so that $\mu(x)=0$, \ie, 
\begin{equation}
  \phi(x)=\theta(x) = \pi\HS(x-x_1) - \pi\HS(x-x_2)
  \label{Eq:FlatState}
\end{equation}
When we look at the temporal evolution of the system we see that such a state is meta-stable and occasionally degrades into two SFs of opposite polarity: a PSF and a NSF. The polarity of a particular SF does not depend on its position and on the direction of $\pi$ jump, but depends on the initial perturbation which we used to disturb the meta-stable state. For example, when we applied a tiny uniform bias current $\gamma=0.01$, we got a NSF at $x=5$ and a PSF at $x=15$. If we apply a bias current of opposite polarity, $\gamma=-0.01$, we get a PSF at $x=5$ and a NSF at $x=15$. This looks natural since, for our choice of current direction, the Lorenz (driving) force associated with the positive current pushes positive (semi-)fluxons to the right and negative ones to the left. Thus, the final configuration always corresponds to a pair of SFs pulled apart by the bias current.

The states of two SFs discussed above, and naturally denoted as \state{du} and \state{ud}, are stable only when the distance between SFs is rather large, $a>a_c$. Otherwise, in the short facet limit, $a<a_c$, the state without SFs is energetically more favorable. The numerical simulations show that the critical separation is $a_c^{(2)}\approx 1.55 \pm 0.05$. The superscript stands for $N_c=2$, as below we also calculate $a_c$ for the case of more than two corners. We have also checked that there is no hysteresis (local energy minimum) around $a_c$, \ie, if we start with two SFs at $a=a_c-\varepsilon$ ($0<\varepsilon\ll1$), the system relaxes to the $\mu=0$ state (\ref{Eq:FlatState}). Inversely, if we start from the flat state (\ref{Eq:FlatState}) at $a=a_c+\varepsilon$, it relaxes to the \state{ud} or \state{du} state.

\begin{figure}
  \centering
  \includegraphics{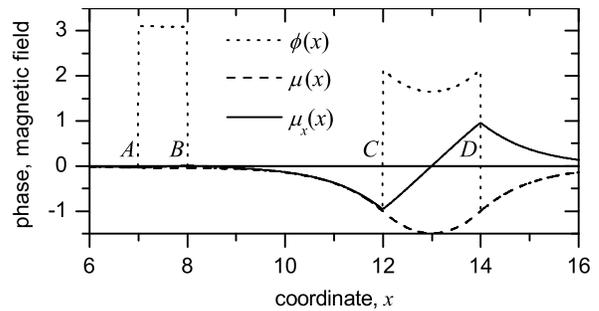}
  \caption{
    Ground state of LJJ with 4 corners, situated in arbitrary way.
  }
  \label{Fig:PSF-NSF-o-pi}
\end{figure}

Continuing our reasoning in a similar fashion, we conclude that for odd $N_c$ in an infinite LJJ, the formation of at least one SF is always energetically profitable. In a LJJ with even $N_c$, there is a crossover distance $a_c$ (which is a function of facet number, size and location) which determines the ground state. 
For equidistantly distributed corners the crossover distances $a_c^{(N_c)}$ were calculated numerically for different number of corners $N_c$. They have the following values: $a_{c}^{(4)}=1.35\pm0.05$, $a_c^{(6)}=1.15\pm0.05$, $a_c^{(8)}=1.05\pm0.05$. For odd $N_c$ the state with $N_c$ AFM-ordered SFs is always favored, \ie, $a_c=0$.
In the general case, the situation can also be mixed, as shown in Fig.~\ref{Fig:PSF-NSF-o-pi}, \ie, the shorter facet $A$--$B$ with $a=1$ has constant $\phi=\pi$, while the longer facet $C$--$D$ with $a=2$ has a PSF-NSF pair at the discontinuity points.

\begin{figure}[\FloatPlacement]
  \centering
  \includegraphics{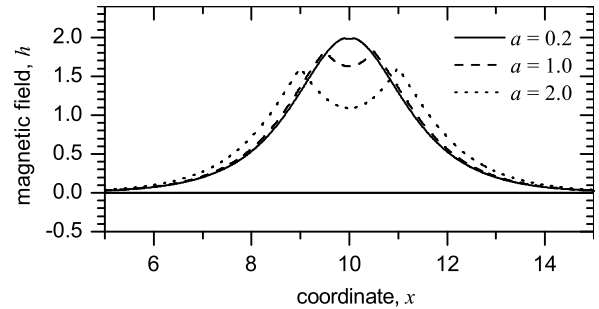}
  \caption{
    The local magnetic field $\mu_x(x)$ of two PSFs for different distance $a=0.1$, $1$, $2$ between them. 
  }  
  \label{Fig:UpUp}
\end{figure}

Another case which we didn't consider up to now is {\bf two PSF} \state{uu} (or two NSF \state{dd}). These states are stable for any separation $a$. For $a \to 0$, the profile of two SFs approaches the profile of a fluxon, as shown in Fig.~\ref{Fig:UpUp}.
The energy of the \state{uu} state changes from $2U_{SF}\approx 4.686$ up to $U_F=8$ as $a$ decreases from infinity to zero. The energy of the \state{ud} state changes from $2U_{SF}\approx 4.686$ down to 0 as $a$ decreases from infinity to zero, having $U(a)\approx a$ for $a\to0$. Although, the energy of the \state{uu} state is always higher than the one of the \state{ud} state, it is still stable and represents a local energy minimum.

\subsection{Finite length LJJ}

We consider here only the case of equidistant distribution of the corners along the length of the LJJ. The $N_c$ corners can be distributed in two different symmetric ways with the positions of the corners: 
\begin{eqnarray}
  x^{(EF)}_k &=& ak              ,\quad a=\frac{L}{N_c+1}
  ; \label{Eq:x1}\\
  x^{(EL)}_k &=& ak -\frac{a}{2} ,\quad a=\frac{L}{N_c}
  , \label{Eq:x2}
\end{eqnarray}
where $a$ is the facet length (the distance between corners), and $k=1 \ldots N_c$. In the first case, all facets have equal length (EF stands for ``Equal Facets''), but for even $N_c$ the total lengths of $0$- and $\pi$-facets are not equal $L_\Sigma^0 \ne L_\Sigma^\pi$. In the second case, the first and the last facets are twice shorter, but the total lengths of 0- and $\pi$-facets are the same for any $N_c$, \ie, $L_\Sigma^0 = L_\Sigma^\pi$ (EL stands for ``Equal (total) Lengths''). The case of one corner\cite{Kirtley::1997:0pi Ic(H),Goldobin:TwinBoundaryComment} is degenerate and belongs to both categories. 

In the analysis below we compare the flat phase state with the antiferromagnetically (AFM) ordered state as they usually represent the lowest energy solutions of Eq.~(\ref{Eq:Final:mu:Norm}). Later we also analyze unipolar states. All other configurations, although stable, are outside the scope of this paper.

Thus, generally, the problem splits into 3 different cases.

(a) \emph{Facets of equal length, even $N_c$, $L_\Sigma^0 \ne L_\Sigma^\pi$.} At $a \gg 1$ the ground state consists of AFM-ordered SFs, since it is not profitable to have long facets with $\phi=\pi$. For estimation of the energy of AFM-ordered state at $a \to 0$, we note that the AFM-ordered SF chain has 
\begin{equation}
  \phi \approx \cos\theta \frac{\pi}{2} = (-1)^l \frac{\pi}{2}
  , \quad a \to 0
  , \label{Eq:PhiApprox}
\end{equation}
where $l$ is the facet number\cite{Next Approx}. Then, at $a \to 0$, the energy of the flat phase state $U_{\mu=0}=N_ca$, while $U_{AFM} \approx L=(N_c+1)a$, \ie,  $U_{\mu=0}<U_{AFM}$. Thus the ground state is a flat phase state. Since we have different limiting behaviors at $a\to0$ and $a \gg 1$, there should be a crossover distance $a_{c}^{(N_c)}$ and a corresponding crossover LJJ length $L_{c}^{(N_c)}$. Our simulations for $N_c=2$ show that $L_{c}^{(2)}=4.25 \pm 0.05$, and $a_{c}^{(2)}=1.4$. For $N_c=4$, $L_{c}^{(4)}=5.85\pm0.05$, and $a_{c}^{(4)}=1.16$. For $N_c=6$, $L_{c}^{(6)}=7$, $a_{c}^{(6)}=1.0$.

(b) \emph{Facets of equal length, odd $N_c$, $L_\Sigma^0 = L_\Sigma^\pi$.} The simplest example $N_c=1$ was already investigated\cite{Kirtley::1997:0pi Ic(H),Goldobin:TwinBoundaryComment}. At $a \gg 1$ the ground state consists of AFM-ordered SFs. At $a\to0$ the flat phase state and the AFM-ordered SF states have the same energy (in the first approximation), so it is reasonable to assume that for all $a$ $U_{\mu=0} \geq U_{AFM}$\cite{U at a=0}, and the ground state is a $\cdots\stat{udu}\cdots$ chain. This is also confirmed by numerical simulations of a $\cdots\stat{udu}\cdots$ SF chain put in LJJs of different lengths.

(c) \emph{Twice shorter edge facets, $L_\Sigma^0 = L_\Sigma^\pi$.} The ground state is equivalent to the previous case and confirmed by simulations.
  
\begin{figure}
  \centering
  \includegraphics{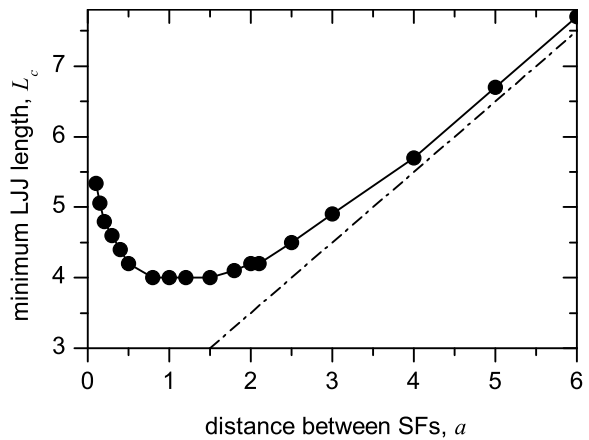}
  \caption{
    The dependence of minimum LJJ length at which the \protect\state{uu} still exists on the distance $a$ between SFs.
  }
  \label{Fig:Lc(a)}
\end{figure}

The last state which we wish to consider is \state{uu} (or \state{dd}). To avoid problems with initial conditions, we have created a \state{uu} state in a LJJ of $L=10$, and have slowly reduced the length of the LJJ in steps of $\delta L=0.1$, allowing the system to relax after each small contraction. We have found that for each separation $a$ there is a critical LJJ length $L_c(a)$, such that at $L<L_c$ two PSFs can not stay inside the junction. They emit a fluxon and revert either to \state{ud} state or to the flat phase state. The plot of $L_c$ {\vs} $a$ is shown in Fig.~\ref{Fig:Lc(a)}.

At $a \to 0$ our two PSFs are equivalent to a fluxon. The pinning disappears too, as a $2\pi$-discontinuity is equivalent to a continuous phase for our problem. A fluxon placed in a LJJ of any finite length is in a metastable state because it is attracted by two images situated behind the left and the right edges of the LJJ. Therefore,
\begin{equation}
  \lim\limits_{a \to 0} L_c(a) = \infty
  , 
\end{equation}
Of course, in the experiment, the force pulling the fluxon towards the boundary is exponentially small for large $L$, and the fluxon gets pinned by non-uniformities or other mechanisms, provided they have a stronger effect than {\eg} thermal fluctuations, which depin the fluxon.

In the opposite limit $a\to\infty$, two PSFs become independent and $L_c \gtrsim a$, \ie, we should solve a problem of SF stability near the edge of a semi-infinite LJJ. This problem, in turn, is equivalent to the problem of stability of a PSF-NSF pair (the NSF appears as an image) in the infinite LJJ analyzed above. Therefore, the SF is stable at the distance from the edge which is larger than $a_c^{(2)}/2$. Thus, in the limit $a\to\infty$, $L_c$ grows linearly with $a$, but is shifted by $a_c^{(2)}$. The asymptote $L_c(a)=a+a_c^{(2)}$ is shown in Fig.~\ref{Fig:Lc(a)} and demonstrates excellent agreement of our arguments with numerical simulation.

\section{Reordering semifluxons by dc bias current}
\label{Sec:Rearrangement}

An interesting question is what happens when a uniform dc bias current is applied to the system. The current acts like a driving force which tries to push SF in a certain direction (just like with fluxons) which depends on the polarity of the SF and on the polarity of the bias current. On the other hand SFs are pinned by the phase discontinuity points and, it seems, the maximum what can happen is that the SF changes the shape (deforms) a little bit. In contrast, we found that the SF can also (a) spontaneously flip and generate a fluxon which starts moving along the LJJ under the action of a bias current, and (b) several PSFs and NSFs can hop simultaneously, if they are situated close enough, resulting in a structural rearrangement.

\subsection{Infinitely long JJ}

First, we simulated a LJJ with only one discontinuity point. According to the results presented in the previous section, the ground state is a SF. Without loosing generality let us assume that it is a PSF. When we apply the positive bias current from zero up to $\gamma_c=0.62$ the PSF just changes shape. At $\gamma=0.63$ the PSF becomes unstable and splits into a positive fluxon and a NSF, so that the total flux is conserved. The fluxon moves away to the right under the action of the driving force. The NSF is still pinned but exposed to the same driving force $\gamma>\gamma_c$ which tries to push it to the left. Correspondingly the NSF splits into a negative fluxon (anti-fluxon) and a PSF. The anti-fluxon moves away to the left being driven by the bias current. Thus, after emission of a fluxon and an anti-fluxon, the system returns to the initial state with one trapped PSF. Since $\gamma>\gamma_c$, the whole process is repeated again, \ie, a SF of any polarity at $\gamma>\gamma_c$ emits two trains of fluxons: fluxons to the right and anti-fluxons to the left.

\begin{figure}[\FloatPlacement]
  \centering
  \includegraphics{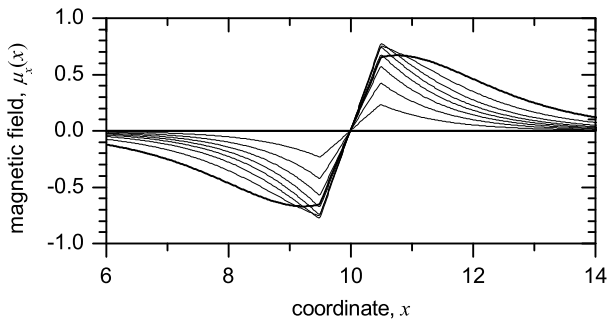}
  \caption{
    The magnetic field $\mu_x(x)$ shows the emerging SFs for $\gamma=0,0.1,\ldots,0.7$ in LJJ of $L=20$ with the distance $a=1$ between corners. The pre-critical state at $\gamma=0.76$ is shown by thicker line.
  }
  \label{Fig:BearingSFs}
\end{figure}

Second, we simulated the behavior of a LJJ with two corners at a distance $a$ between them. When the distance $a<a_c^{(2)}$ (we took $a=1$ for simulations), the ground state corresponds to the flat phase profile. Nevertheless, if one applies a bias current, a NSF--PSF pair is formed. The SFs of this pair are pinned at the corresponding discontinuity points and the polarity of the SFs is such that the bias current pulls them apart. The amplitude (the field at the center of the PSF) of the NSF--PSF pair grows smoothly with the applied bias current and reaches the value $h\approx0.75$ at $\gamma_c=0.77$, as shown in Fig.~\ref{Fig:BearingSFs}. 
This value of field is only twice smaller than the field in the center of an isolated SF, which is equal to $\sqrt{2}\approx1.4$. This means that \emph{even in the short facet limit a NSF--PSF pair can be observed experimentally\cite{SQUID-mu-scope} by applying a uniform dc bias current through the junction}. At $\gamma>\gamma_c$, the NSF--PSF pair emits fluxons to the right and anti-fluxons to the left (for positive bias current), and, thus, switches to the non-zero voltage state.

Another possibility is that the distance between two corners $a$ is larger than $a_c^{(2)}$ (we took $a=2$ for simulations). In this case the ground state can be either of \state{dd}, \state{du}, \state{ud}, \state{uu}, depending on the history of the system. Under the action of the applied bias current, the states \state{dd} and \state{uu} behave very similar to a single SF state discussed above. The most interesting thing happens when we apply the bias current to the \state{du} or \state{ud} configuration, especially if the initial order of NSF and PSF is not natural with respect to the direction of bias current. Suppose that we have the state \state{ud}. For $\gamma>0$ the direction of driving force is such that the NSF and the PSF are pushed towards each other, but still they can not move since they are pinned at the corners. Only their shape changes a little bit as $\gamma$ increases. At $\gamma=0.08$, we observe that the SFs flip synchronously, and exchange their location. We can denote such a process as 
\state{ud}$\stackrel{\gamma=0.08}{\longrightarrow}$\state{du}.
This process can be thought as an exchange by a \emph{virtual} fluxon which transfers a quantum of magnetic flux from the PSF to the NSF:
\[
  \mbox{PSF$\longrightarrow$NSF + Virtual Fluxon}
\]
\[
  \mbox{Virtual Fluxon + NSF$\longrightarrow$PSF}
\]
We use the word \emph{virtual} because (a) no actual soliton is formed (only a quantum of magnetic flux spills over), and (b) this happens at the value of $\gamma$ which is much less than the fluxon creation current $\gamma_c=0.63$ discussed above. The process of flipping (or hopping PSF$\rightleftarrows$NSF) of SFs is accompanied by emission of plasma waves in both directions, but they decay with time, so that the final state \state{du} is stable. In this new state the bias current pulls the SFs apart and only when $\gamma$ exceeds $\gamma_c=0.67$ the system switches to non-zero voltage state. If the system was initially in the state \state{du}, no structural rearrangement takes place for $\gamma=0\ldots\gamma_c=0.67$ and the system switches to the resistive state at $\gamma=\gamma_c$. Thus, \emph{by sweeping the bias current back and forth between $-\gamma_c$ and $+\gamma_c$ one may observe\cite{SQUID-mu-scope} the structural rearrangements of SFs being in zero voltage state}.

Similar rearrangements we observe for the states involving many SFs. Usually we start from an array of AFM-ordered SFs at $\gamma=0$ with the distance $a=2$ between nearest neighbors. Then we increase the bias current to obtain the following structural transformations.

  For 4 SFs: 
  \begin{eqnarray}
    &&
    \stat{udud}\stackrel{\gamma=0.12}{\longrightarrow}
    \stat{dudu}\stackrel{\gamma=0.25}{\longrightarrow}\stat{dduu}
  \end{eqnarray}
  with $\gamma_c=0.53$ from the state \state{dduu}.
    
  For 6 SFs:
  \begin{eqnarray}
    &&
    \stat{ududud}\stackrel{\gamma=+0.13}{\longrightarrow}
    \stat{dduduu}\stackrel{\gamma=+0.30}{\longrightarrow}
    \stat{ddduuu}\stackrel{\gamma=+0.13}{\longrightarrow}  
    \nonumber\\&&
    \stat{dduduu}\stackrel{\gamma=-0.03}{\longrightarrow}
    \stat{ududud}\stackrel{\gamma=-0.21}{\longrightarrow}
    \stat{uuuddd}
  \end{eqnarray}
  with $\gamma_c=0.50$ from the state \state{ddduuu}.
  In the latter case the bias was first increased and then decreased, thus demonstrating a hysteresis in switching between various configurations. 
  
  Further, for 8 SFs: 
  \begin{eqnarray}
    &&\stat{udududud}
    \stackrel{\gamma=+0.13}{\longrightarrow}\stat{ddududuu}
    \stackrel{\gamma=+0.23}{\longrightarrow}\stat{dddduuuu}
    \stackrel{\gamma=+0.14}{\longrightarrow}
    \nonumber\\
    &&\stat{ddduduuu}
    \stackrel{\gamma=-0.02}{\longrightarrow}\stat{udududud}
    \stackrel{\gamma=-0.20}{\longrightarrow}\stat{uuuudddd}
  \end{eqnarray}
  with $\gamma_c=+0.49$ from \state{dddduuuu} state. 
  Note, that sometimes the system switches directly to the very polarized state skipping several intermediate configurations. Those intermediate states can sometimes be obtained by sweeping the bias current back and forth, starting from various other states. The state \state{ddduduuu} in the above example was obtained so.
  
  For 12 SFs: 
  \begin{eqnarray}
    &&
    \stat{udududududud}\stackrel{\gamma=+0.14}{\longrightarrow}
    \stat{ddddduduuuuu}\stackrel{\gamma=+0.31}{\longrightarrow}
    \nonumber\\&&
    \stat{dddddduuuuuu}\stackrel{\gamma=+0.12}{\longrightarrow}
    \stat{ddddduduuuuu}\stackrel{\gamma=+0.00}{\longrightarrow}
    \nonumber\\&&
    \stat{ddudduduuduu}\stackrel{\gamma=-0.05}{\longrightarrow}
    \stat{uudududududd}\stackrel{\gamma=-0.18}{\longrightarrow}
    \nonumber\\&&  
    \stat{uuuuududdddd}\stackrel{\gamma=-0.30}{\longrightarrow}
    \stat{uuuuuudddddd}
    \label{Eq:12SFs}
  \end{eqnarray}
  with $\gamma_c=+0.47$ from the state \state{dddddduuuuuu}. We also checked that the situation is symmetric, \ie, $\gamma_c=-0.47$ from the state \state{uuuuuudddddd}. If we start from the state \state{ddudduduuduu} at $\gamma=0$ obtained in (\ref{Eq:12SFs}), and increase $\gamma$, we get some new states which we have not seen before:
  \[
   \stat{ddudduduuduu}\stackrel{\gamma=+0.12}{\longrightarrow}
   \stat{dddudududuuu}\stackrel{\gamma=+0.19}{\longrightarrow}
   \stat{dddddduuuuuu}
  \]

\begin{figure}[\FloatPlacement]
    \centering 
    \includegraphics{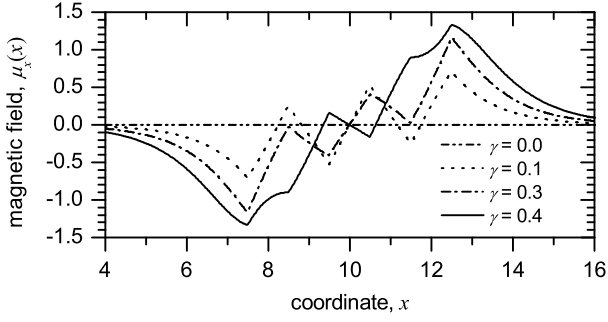}
    \caption{
       The magnetic field $\mu_x(x)$ profiles for different configurations obtained during transformation (\ref{Eq:Emerge&Arrange}) for $\gamma=0, 0.3, 0.4$.
    }
    \label{Fig:Emerge&Arrange}
    \end{figure}

  Interesting enough, similar rearrangements are possible even when the initial state at $\gamma=0$ is flat. For example, for 6 SFs separated by the distance $a=1$ we get
  \begin{eqnarray}
    \texttt{------}\stackrel{\gamma>+0.00}{\longrightarrow}
    \stat{dududu}\stackrel{\gamma=+0.37}{\longrightarrow}
    \stat{dd}\texttt{--}\stat{uu}  
    \label{Eq:Emerge&Arrange}
  \end{eqnarray}
  with $\gamma_c=0.48$ from the state \state{dd}\texttt{--}\state{uu}, see Fig.~\ref{Fig:Emerge&Arrange}.
    %

\subsection{LJJ of finite length}

Is it possible to observe similar ``polarization by current'' effects when the SFs are equidistantly distributed along a LJJ of finite length? For the annular LJJ the answer should be negative, as, due to symmetry, there are no left and right edges. On the other hand, for a linear LJJ, it seems possible. As before we distinguish 3 different cases.
  
(a) \emph{Facets of equal length, even $N_c$.} The ground state at $\gamma=0$ can be flat for $a<a_{c}^{N_c}$ or consist of AFM-ordered SFs for $a>a_{c}^{N_c}$. Let us start with the AFM-ordered state \state{ud} at $L=6$ ($a$=2). We observe the transition 
\begin{equation}
  \stat{ud}\stackrel{\gamma=+0.083}{\longrightarrow}\stat{du}
  , 
\end{equation}
with $\gamma_c=0.491$ from the \state{du} state.
  
For $N_c=4$:
\begin{equation}
  \stat{udud}\stackrel{\gamma=+0.119}{\longrightarrow}\stat{dduu}
  , 
\end{equation}
with $\gamma_c=0.331$ from the state \state{dduu}.

For $N_c=6$:
\begin{equation}
  \stat{ududud}\stackrel{\gamma=+0.12}{\longrightarrow}\stat{dduduu}
  , 
\end{equation}
with $\gamma_c=0.28$ from the state \state{dduduu}. As we see in the latter case the total polarization does not take place.

For $N_c=8$:
\begin{eqnarray}
  &&\stat{udududud}
  \stackrel{\gamma=+0.14}{\longrightarrow}\stat{ddududuu}
  \stackrel{\gamma=+0.23}{\longrightarrow}\stat{dddduuuu}
  \stackrel{\gamma=+0.14}{\longrightarrow}
  \nonumber\\
  &&\stat{ddduduuu}
  \stackrel{\gamma=-0.01}{\longrightarrow}\stat{dudududu}
  \ldots
\end{eqnarray}
with $\gamma_c=0.27$ from the state \state{dddduuuu}. It is rather interesting that the rearrangements happen almost at the same value of $\gamma$ as in the case of an infinitely long LJJ. We also have noticed that as $a$ decreases towards $a_c^{N_c}$ or as $N_c$ increases at fixed $a$, the current $\gamma^*$ at which the first rearrangement takes place also decreases, reaching $\gamma^*=0$ at $a \to a_c^{N_c}$. In this limit the total polarization is almost never achieved.

When $a<a_{c}^{N_c}$ the initial state is flat at $\gamma=0$, but AFM-ordered SFs emerge at $\gamma>0$. The amplitude of magnetic field at the centers of SFs grows with $\gamma$. Such states can not be rearranged further and usually the system switches to the resistive state at some $\gamma_c$. The results are summarized in Tab.~\ref{Tab:EqFacets-EvenNc-SmallA}. The simulations also show that for $a<a_{c}^{N_c}$ the system is kind of ``elastic', \ie, as soon as the force $\gamma$ is removed, the system, like a spring, returns back to the flat phase state without any hysteresis.
\begin{table}
  \begin{tabular}{  d  d  d  }
    a & N_c & \gamma_c\\
    \hline\hline
    0.5 & 2 & 0.33\\
    0.5 & 4 & 0.20\\
    0.5 & 6 & 0.14\\
    0.5 & 8 & 0.11\\
    \hline
    1.0 & 2 & 0.36\\
    1.0 & 4 & 0.23\\
    1.0 & 6 & 0.17\\
    1.0 & 8 & 0.14\\
    \hline
    1.3 & 4 & 0.11\\
    1.3 & 6 & 0.18\\
    \hline\hline
  \end{tabular}
  \caption{
    The list of numerical tests which shows the $\gamma_c$ for the switching from the \protect\state{du}-like states. No structural rearrangements were observed (even $N_c$).
  }
  \label{Tab:EqFacets-EvenNc-SmallA}
\end{table}

(b) \emph{Facets of equal length, odd $N_c$.} The ground state at $\gamma=0$ is an AFM-ordered $\ldots\stat{udu}\ldots$ chain, so we start with our usual value $a=2$. The smallest odd $N_c$ for which it makes sense to speak about rearrangements is \state{udu}:
\begin{equation}
  \stat{udu}\stackrel{\gamma=+0.17}{\longrightarrow}\stat{duu}
  , 
\end{equation}
with $\gamma_c=0.32$ from the state \state{duu}.

For $N_c=5$:
\begin{equation}
  \stat{ududu}\stackrel{\gamma=+0.14}{\longrightarrow}
  \stat{duduu}\stackrel{\gamma=+0.27}{\longrightarrow}
  \stat{dduuu}    
  , 
\end{equation}
with $\gamma_c=0.28$.

\begin{figure}[\FloatPlacement]
  \centering
  \includegraphics{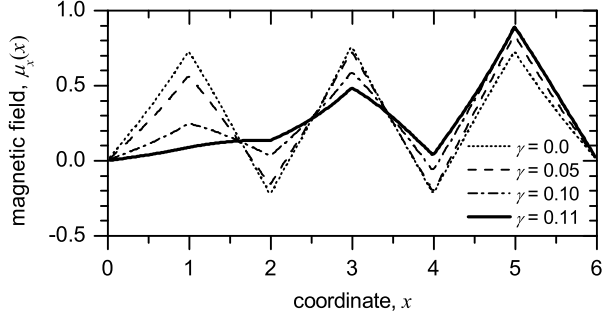}
  \caption{
    The magnetic field $\mu_x(x)$ profiles for different values of bias current $\gamma=0, 0.05, 0.10, 0.11$ in LJJ of $L=6$ with $N_c=5$ corners and equal facet lengths. The field of SFs is ``pulled'' to the right by the bias current.
  }
  \label{Fig:PullField}
\end{figure}
For small $a$ (we have used $a=1$) the rearrangement of SFs does not take place, although the amplitude of magnetic field in the center of each SF changes so that the amplitude of field is higher at the right edge and lower at the left edge. It is like the field is globally pulled toward the right edge by the bias current, as can be seen in Fig.~\ref{Fig:PullField}
\begin{table}
  \begin{tabular}{  d  d  d  l  }
    a & N_c & \gamma_c & rearrangements\\
    \hline\hline
    0.5 & 3 & 0.04\\
    0.5 & 5 & 0.04\\
    0.5 & 7 & 0.04\\
    \hline
    1.0 & 3 & 0.13\\
    1.0 & 5 & 0.10\\
    1.0 & 7 & 0.08 & inv. left SFs\\
    \hline
    1.5 & 3 & 0.17 &
    \state{udu}$\stackrel{\gamma=+0.16}{\longrightarrow}$\state{duu}\\
    1.5 & 5 & 0.13 & 
    \state{ududu}$\stackrel{\gamma=+0.14}{\longrightarrow}$\state{duduu}\\
    1.5 & 7 & 0.13 & \state{udududu}$\stackrel{\gamma=+0.08}{\longrightarrow}$\state{dududuu}\\
    \hline\hline
  \end{tabular}
  \caption{
    The list of numerical tests which shows the $\gamma_c$ for the switching from the \protect\state{du}-like states in LJJ with uniformly distributed corners and equal facet lengths. No structural rearrangements were observed (odd $N_c$).
  }
  \label{Tab:EqFacets-OddNc-SmallA}
\end{table}
The results of other runs for this case are summarized in Tab.~\ref{Tab:EqFacets-OddNc-SmallA}.

(c) \emph{Twice shorter edge facets.} The ground state at $\gamma=0$ consists of AFM-ordered SFs. Numerical results show that no rearrangements were observed for $N_c=8,\ldots,18$ in a LJJ of $L=30$ ($a$ from $3.75$ to $1.67$). For smaller $a$ rearrangements were observed in few cases at $a\to0$, see Tab.~\ref{Tab:HighSFDensity}. Such a difficulty of rearrangement can actually be explained. If we remove the LJJ edges and introduce SF images, our problem reduces to the problem of rearranging an infinite chain of AFM-ordered SFs. Similar to the case of an annular LJJ, there is no left and right ends here and all points are equivalent, so the rearrangement is never initiated. Synchronous hopping of all SFs would require infinite energy. Of course, in an experiment the facets are never perfectly equal, so this may result in rearrangements of SFs. In simulations rearrangements may take place due to the finite $\Delta x$ of the numerical scheme or the finite step in $\gamma$. At $a \to 0$ the rearrangement, if any, happens at $\gamma \to 0$.
\begin{table}
  \begin{center}
    \begin{tabular}{  d  d  d  l }
      a  & N_c & \gamma_c & rearrangements\\
      \hline\hline
      2.0  & 2 & 0.15  & none\\
      1.0  & 2 & 0.038 & none\\
      0.75 & 2 & 0.020 & none\\
      0.5  & 2 & 0.020 & 
      $\stat{ud}\stackrel{\gamma=+0.005}{\longrightarrow}\stat{du}$\\
      \hline
      2.0  & 3 & 0.155 & none\\
      1.5  & 3 & 0.094 & none\\
      1.0  & 3 & 0.044 & none\\
      0.5  & 3 & 0.015 & none\\
      \hline
      2.0  & 4 & 0.15  & none\\
      1.5  & 4 & 0.091 & none\\
      1.0  & 4 & 0.040 & none\\
      0.5  & 4 & 0.014 & 
      $\stat{udud}\stackrel{\gamma=+.008}{\longrightarrow}\stat{dudu}$\\
      \hline
      2.0  & 5 & 0.15  & none\\
      1.5  & 5 & 0.093 & none\\
      1.0  & 5 & 0.043 & none\\
      0.5  & 5 & 0.014 & none\\
      \hline
      2.0  & 6 & 0.15  & none\\
      1.5  & 6 & 0.091 & none\\
      1.0  & 6 & 0.040 & none\\
      0.5  & 6 & 0.013 & $\stat{ududud}
      \stackrel{\gamma=+.009}{\longrightarrow}\stat{dududu}$\\
      \hline
      2.0  & 7 & 0.15 & none\\
      1.5  & 7 & 0.092 & none\\
      1.0  & 7 & 0.042 & none\\
      0.5  & 7 & 0.012 & none\\
      \hline
      2.0  & 7 & 0. & none\\
      1.5  & 7 & 0. & none\\
      1.0  & 7 & 0. & none\\
      0.5  & 7 & 0.012 & $\stat{udududud}
      \stackrel{\gamma=+.009}{\longrightarrow}\stat{dudududu}$\\
      \hline
      \hline
    \end{tabular}
  \end{center}
  \caption{
    The list of numerical tests which shows the value of $\gamma_c$ for the switching from the \protect\state{du}-like states in a LJJ with uniformly distributed corners and twice shorter edge facets. No structural rearrangements were observed.
  }
  \label{Tab:HighSFDensity}
\end{table}

The hopping of SFs under the action of bias current results in a possibility to observe half-integer Zero-Field Steps (ZFS) in a LJJ with moderate facet size\cite{Stefanakis:ZFS/2}. If we take a LJJ of length $L=2$ with a phase discontinuity point and a PSF in the middle, the interaction of the PSF with the boundaries at zero applied magnetic field can be treated by introduction of two NSF (images) at the distance 1 from the left and right edge of the LJJ. In this case the boundary can be removed and we have a pinned NSF-PSF-NSF system with $a=2$ in an infinite LJJ. Under the action of a bias current $\gamma>\gamma^*$, the PSF exchanges its position with one of the images (NSF) so that we get a NSF in the center of LJJ and one quantum of magnetic flux (virtual fluxon) transferred through the edge. Now the system is equivalent to PSF-NSF-PSF with $a=2$ in an infinite LJJ and the whole process is repeated again, transferring another virtual fluxon through the edge of the LJJ. At this point the state of the system is exactly equal to the initial state and the whole period is repeated again. By analogy with ZFS, the PSF is also ``reflected'' as an NSF after interaction with the edge, but this process is not continuous as in the case of fluxons, but somewhat discrete --- a PSF hops out and a NSF hops in. Since the total flux transferred per ``reflection'' is $2\pi$ (twice smaller than for conventional ZFSs), and the velocity of the virtual fluxon presumably does not exceed the Swihart velocity one should observe half-integer ZFS in the experiment.

\section{Conclusions}
\label{Sec:Conclusion}

We have discussed various configurations of semifluxons (SFs) in long Josephson junctions (LJJs) with $\pi$-discontinuities (zigzag corners) and corresponding ground states. We have shown that 
(a) in an infinite LJJ the ground state depends on the separation between the corners (facet size), and may be either a flat phase state, or an array of antiferromagnetically ordered SFs, or an array of unipolar SFs. If the number of corners is odd, at least one SF should be present.
(b) In a LJJ of finite length the ground state of a LJJ with $N_c$ uniformly distributed corners can be flat or consist of SFs depending not only on the facet size $a$, but also on the question whether $N_c$ is an odd or even number and on the way to distribute the corners.
(c) In the case of a flat phase ground state, one may still compel SFs to emerge by applying a uniform dc bias current.
(d) The rearrangements of SFs by bias current, discussed in this paper, can be observed experimentally\cite{SQUID-mu-scope} by being in the zero-voltage state and changing the current in the range from $-\gamma_c$ to $+\gamma_c$. One may observe the whole variety of states discussed here, if the number of facets and their sizes are properly selected.
(e) Although the SFs are pinned, they may hop from one discontinuity point to the next one, provided the distance between these points is about $2\lambda_J$ or less. This "moving by hopping" results in a possibility to observe some phenomena known from fluxon dynamics, \eg, zero-field steps, but with the first step situated at half of the voltage of the usual first zero-field step.

The most numerical observations reported here can be directly checked in a SQUID microscopy experiment. The results presented here provide new insights into the statics and \emph{dynamics} of SFs in LJJs with $\pi$-discontinuities. We also hope that this work will stimulate further research and generate ideas for novel devices and their application in superconducting electronics.

\begin{acknowledgments}
  We would like to thank H.~Hilgenkamp, H.-J. Smilde, N.~Stefanakis and C.~C.~Tsuei for interesting discussions.
\end{acknowledgments}


\begin{thebibliography}{99}
\bibitem{Goldobin:0Pi:AlphaEstimation}  
  Actually the estimation of $\alpha$ is tricky, and will be discussed in detail elsewhere.
\bibitem{Next Approx} 
  The next approximation would be to assume that 
  $\phi\approx\frac{\pi}{2}\cos\theta + \varepsilon(x)$
\bibitem{U at a=0}  
  Using the next order approximation it can be shown that $U_{AFM}\approx a - c a^3$ ($c>0$), while $U_{\mu=0} = a$, \ie, AFM SF state has lower energy at any $a$.
\end{thebibliography}
\end{document}